\documentclass[notitlepage]{article}
\usepackage{titling}
\usepackage{lipsum}

\usepackage{epsfig}
\usepackage{color,soul}
\usepackage{graphics}
\usepackage[normalem]{ulem}
\usepackage{epsfig}
\usepackage{cite}
\usepackage{amsmath}
\usepackage{mathtools}
\usepackage{textgreek}
\usepackage[dvipsnames]{xcolor}
\definecolor{noble}{RGB}{144, 18, 175}
\title{
Tidal deformability of neutron stars with exotic particles within a density dependent relativistic mean field model in R-squared gravity}

\author{K Nobleson, Tuhin Malik,  and Sarmistha Banik}

\date{\today}
\begin{document}

\maketitle
\thispagestyle{empty}

%sarmistha.banik@hyderabad.bits-pilani.ac.in
          % Insert a name or remove this line
%
%\institute{Birla Institute of Technology and Science Pilani, Hyderabad Campus, Hyderabad - 500078, India }
\begin{abstract}
There is a growing interest in investigating modified theories of gravity, primarily, with the aim of explaining the universe's accelerated expansion, which has been confirmed by several independent observations. Compact objects, like neutron stars, exhibit strong gravity effects and therefore are used to study modified gravity theories. We use the $f(R)=R+aR^2$ model, where R is the Ricci scalar and $a$ is a free parameter. This model has been studied both perturbatively and non-perturbatively. However, it was found that perturbative methods results in nonphysical solutions for the neutron stars. In this paper, we examine neutron star properties, such as mass, radius, tidal deformability in non-perturbative $f(R)$ gravity model with density dependent relativistic equation of state with different particle compositions. The strange particles in the core of NS in the form of ${\bf \Lambda}$ hyperons, $K^-$ condensate, and quarks are considered. We have observed that while the mass-radius relation allows for a wide range of parameter $a$, when tidal deformability is considered, the parameter $a$ is constrained down by one order.
\end{abstract}
%
%\PACS{
%      {PACS-key}{discribing text of that key}   \and
%      {PACS-key}{discribing text of that key}
%     } % end of PACS codes
 %end of abstract
%
%\maketitle
%\onecolumn
\section{Introduction}
\label{intro}

Einstein's General Theory of Relativity (GR)\cite{GR} has been the most influential theory of the twentieth century, providing the most logical explanation of the Universe at large scales. It, along with the Standard Model of particle physics, constitutes the "Grand Unified Theory" from which all physics can be derived. The most important insight of GR is that it defines gravity, one of the fundamental forces, not as a force but as an intrinsic property of space-time that results from viewing the Universe as a geometric manifold. The conceptual complications in quantizing Einstein's theory, as well as, the astrophysical findings of accelerated expansion of universe, indicated that GR might need to be modified. 

In the recent years, there has been a lot of interest in studying modified theories of gravity, primarily with the intention of finding physical explanations for the universe's accelerated expansion, which has been validated by many independent observations \cite{perlmutter1999measurements}. The cosmological constant is the simplest solution that can be made consistent with recent observations of cosmic acceleration. However, the field theoretic understanding of cosmological constant is far from satisfactory, and its magnitude is significantly smaller than predicted, resulting in the well-known {\it coincidence} and {\it fine-tuning} problems. To address this problem, many approaches have been proposed in the literature \cite{tsujikawa2006dynamics}. The two most popular approaches that have been used are $-$ either introducing an additional scalar field  or replacing the Ricci scalar $R$ in Einstein-Hilbert action with a general function of the Ricci scalar $f(R)$ \cite{carroll2004cosmic}. The additional scalar degree of freedom in both formulations can be configured to simulate the cosmological constant or any type of plausible cosmological evolution at any cosmological scale. The functional form of $f(R) = R +aR^2$ is used in this study. The modified theories of gravity are well-motivated and compatible with observational evidence without the requirement of the addition of a previously undetected component of matter. Such modifications, on the other hand, should be subjected to all the astrophysical tests before being approved or dismissed. These models are verified for their feasibility in the weak gravity regime by the laboratory tests \cite{ozel2016masses} and solar system tests. However, any coherent theory of gravity, whether classical or modified, should adhere to the strong gravity regime as well. Compact stars, such as neutron stars, are ideal for studying strong gravity, high-energy nuclear physics, QCD under extreme conditions, and putting alternate theories of gravity to the test, among other things.
%Compact stars, such as neutron stars, provide an excellent platform for investigating strong-gravity behaviour, high-energy nuclear physics, QCD under extreme conditions, and for testing alternative theories of gravity, etc. 

In general, the equations in the $f(R)$ model can be solved using two approaches: a) perturbative approach, b) non-perturbative approach. In perturbative approach, the $f(R)$ theories are considered as perturbation to GR and the next to the leading order terms in a larger expansion are used to treat the difficulty of field equations \cite{Arapo_lu_2011,  deliduman2012neutron, orellana2013structure, Astashenok}. In non-perturbative approach, the equations are solved self-consistently by solving the interior and exterior equations with appropriate boundary conditions \cite{Daniela1, plb}. Since, in the perturbative approach, the interior and exterior solutions are not solved simultaneously and the allowed values of $a$ can be negative, it was argued in \cite{Daniela1} that solutions in this method would yield non-physical results for the NS. In order to avoid tachyonic instabilities and non-physical results, we use non-perturbative method to examine the existence and properties of neutron star (NS) with realistic equation of state (EoS) in the context of the $f(R)$ model, and obtain the possible constraints on the parameter $a$. 

The nature of high density matter in the interior of NS is still a subject of active research. The density of NS exceeds the supra-nuclear density ($\rho_0\sim 2.4 \times 10^{17}kg/m^3$) at the core. However, due to our limited understanding beyond the saturation density, the high-density existence of nuclear matter remains unknown. There are several models for constructing the EoS at these densities that are then modified based on nuclear experimental findings and astrophysical observations. Usually two types of treatment exist for constructing EoS i) Relativistic and ii) Non-relativistic. Although, non relativistic method have been extremely successful in the description of nucleons inside atomic nuclei (finite nuclei) but for dense nuclear matter study relevant for NS, the choice of relativistic effect is more appropriate. The force or the interactions between two nucleons is understood as the exchange of virtual particles known as mesons which transfer momentum and energy between the two nucleons. In order to exchange them, the nucleons must be extremely close together of the order of 1 Fermi ($10^{-15} m$), beyond which the strong nuclear force decreases sharply and becomes almost negligible. This is the relativistic description of nuclear interaction and is the basis of all successful models of nuclear structure. In this case, the nuclear matter is described by a Lagrangian density depicting a many-body system. The first field-theoretic attempt to derive the nucleon-nucleon (NN) interaction via pion-exchange was carried out by Yukawa \cite{Yukawa:1935xg} where baryons were interacting with a neutral scalar field and it was extended by proposing a neutral vector field \cite{Proca:1900nv}. The empirical values of nuclear matter properties at normal nuclear matter density, specifically, the binding energy, symmetry energy and its slope, effective mass of nucleons, compressibility constrain the relativistic EoSs of dense matter. Among the relativistic models, the DD2 \cite{hempel2010statistical, char2014massive} is constructed following the Walecka prescription \cite{wal} and is one of the successful models. 
The interesting features of this kind of model is the  high-density behaviour is considered through the density dependence of the meson-baryon couplings. Later, this DD2 model extended nucleons with additional strange components \cite{banik2014new, apj2021, oosA}.

In the absence of direct evidence, the only reliable observables are its mass and radius measurements. Since the GW170817 detection of the merger of two orbiting NS \cite{abbott2017gw170817,abbott2018gw170817}, additional parameters like tidal deformability ($\Lambda$) \cite{Tanja} is now available to study the interior of the NS. The high-density nucleus of a compact star is believed to have more degrees of freedom than neutrons, protons, and electrons. The EoS softens in the presence of exotic particles, resulting in an overall lower mass NS. Therefore, EoS can be constrained by observational mass. The mass measurements of two massive pulsars, with masses of $1.928\pm 0.017 M_{\odot}$ \cite{fonseca2016nanograv} and $2.01\pm0.04 M_{\odot}$ \cite{antoniadis2013t}, ruled out the soft EoS, which are incapable of supporting such a massive star. However, a few exotic EoS comprising hyperons, antikaons, and quarks are consistent with the above findings. GW170817, the first observation of gravitational waves from a binary NS merger, has provided an upper limit on tidal deformability \cite{abbott2017gw170817, abbott2018gw170817}, which has been used by many authors to constrain the NS radius. A comprehensive list can be found in Table 1 of the review \cite{Luca}.  
A constraint of $9.8 < R_{1.4} < 13.2$ km is given in Ref \cite{raithel2019constraints}, which is consistent with the radius measurement obtained from X-ray observations. Mass and size data for the pulsar J0030+0451 \cite{2019ApJ...887L..22R} have been published by NASA's Neutron star Interior Composition Explorer (NICER). According to these studies \cite{2019ApJ...887L..22R, Miller_2019}, it has a mass of $1.3-1.4M_{\odot}$ and a radius of $\sim 12.7-13 $ km. 

The dimensionless tidal deformability parameter can be used to decode the gravitational wave phase evolution induced by the deformation. It is a measure of the neutron star's response to gravitational pull on the surface that correlates with the pressure gradient within the NS and is highly dependent on the NS internal structure. Many  authors  investigated the  rich connection  between  tidal deformability and 
the dense matter EoS more intensely after the detection of gravitational waves from the GW170817 event \cite{DB, PChar1, de, Radice}. To investigate dense matter EoS, precise measurements of the $\Lambda$ and radius of NS can be used as an effective probe \cite{Tmalik}. 

Given an EoS, the physical properties, such as mass-radius of a NS are obtained by solving the Tolman-Oppenheimer-Volkoff (TOV) equations. In $f(R)$ formulation, a spherically symmetric, non-rotating mass distribution under hydrostatic equilibrium can be described by a set of modified TOV equations. In scalar tensor theories of gravity these modified TOV solutions reproduce the correct behaviour at the weak gravity limit as shown in \cite{babichev2010relativistic}. Realistic EoS of the density-dependent models described above and the modified TOV equations are used to derive the measurable properties of NS in this study.

Currently, there is an overlap in the uncertainties arising due to our lack of complete understanding of the EoS and the free parameter of the modified gravity theories. A strong constraint of the free parameter in the modified gravity theory will improve our understanding of the EoS and vice-versa. In an earlier work by \cite{Daniela1}, they used EoS, such as, SLY4 and APR4 to calculate the mass, radius, tidal love number ($k_2$), in non-perturbative $f(R)$. In this study, we investigate this further with EoS with different constituents, such as, quarks, strange particles, and calculate the mass and radius for different values of $a$ and place an upper bound on $a$ based on the observational constraints. Next, we calculate the tidal deformability and constrain the value of $a$ further based on the observational constraint from GW170817 event. 

The following is the breakdown of the paper's structure. In section II, we use the non-perturbative formalism to study the field equations of the $f(R)$ gravity model. The modified TOV equations are derived from the non-perturbative forms of metric functions and hydrodynamical quantities. In section III, we present an overview of the EoS used in this work and the results of the numerical solution of mass, radius, tidal love number, and tidal deformability for various forms of the EoS, and various values of the free parameter $a$. Finally, in the conclusion section, we comment on the numerical results as well as the constraint we obtained for the free parameter $a$ of the $f(R)$ gravity model.

\section{\label{sec:mTOV} $f(R)$ Gravity and the Modified TOV Equations}
The $4$-dimensional action of $f(R)$ gravity is the simplest generalization of the Lagrangian in the Einstein-Hilbert action as \cite{Sotiriou}:
\begin{eqnarray}\label{action}
S=\frac{1}{16\pi}\int d^4x \sqrt{-g}f(R) + S_{{\rm m}}
\end{eqnarray}
where $g$ denotes the determinant of the metric $g_{\mu\nu}$, $R$ is the Ricci scalar with respect to the spacetime metric $g_{\mu\nu}$, and $ S_m$ is a matter action. We have set $G$ and $c$ to $1$ in the action. 
The $f(R)$ theories are made free of tachyonic instabilities, appearance of ghosts, and made viable by imposing \cite{Sotiriou, De_Felice}
\begin{eqnarray}\label{ghosts}
\frac{df}{dR} > 0; \quad \frac{d^2f}{dR^2} \geq 0
\end{eqnarray}
The $f(R)$ theories are well established to be analogous to the Brans-Dicke scalar-tensor theory with $\omega_{BD} = 0$ and a  potential for the scalar field. The dynamically equivalent action is given by this new field $\psi$ as
\begin{eqnarray}\label{dyn_action}
S=\frac{1}{16\pi}\int d^4x \sqrt{-g}[f(\psi)+f'(\psi)(R-\psi)] + S_{{\rm m}}
\end{eqnarray}
Varying the action with respect to $\psi$, we obtain $f''(\psi)(R-\psi) = 0$, which gives $\psi = R$ if $f''(\psi) = 0$. By introducing a new field $\Phi =f'(\psi) $ and defining the potential $U(\Phi) $ as 
\begin{eqnarray}\label{poten}
U(\Phi) = \psi(\Phi)f'(\psi(\Phi)) - f(\psi(\Phi)) 
\end{eqnarray}
will take the action in eqn \ref{dyn_action} exactly to the form of action of Brans-Dicke Theory in Jordan frame with a potential for the scalar field as \cite{Daniela2}
\begin{eqnarray}\label{dyn_action_pot}
S=\frac{1}{16\pi}\int d^4x \sqrt{-g}[\Phi R- U(\Phi)] + S_{{\rm m}}
\end{eqnarray}
In the case of R-squared gravity, the Brans-Dicke potential is given by
\begin{equation}\label{BD_pot}
U(\Phi) = \frac{1}{4a}(\Phi- 1)^2 
\end{equation}
which corresponds to a massive scalar field with mass $m_{\Phi} = 1/\sqrt{6a}$. Here we consider only the values of $a$ which satisfies the condition $ d^2f/ dR^2 \geq 0 $. From a mathematical and numerical perspective, it is more convenient to study the field equations in the Einstein frame. An Einstein frame can be defined by the introduction of a new scalar field $\varphi$,  
\begin{eqnarray}\label{phi}
\varphi = \frac{\sqrt{3}}{2}ln\Phi 
\end{eqnarray}
and the new metric $g^*_{\mu \nu}$ is given by
\begin{eqnarray}\label{BD_pot2}
g^*_{\mu \nu} = \Phi g_{\mu \nu} = A^{-2}(\varphi) g_{\mu \nu} 
\end{eqnarray}
with $A^2(\varphi) = \Phi^{-1}(\varphi) = exp(-2\varphi/\sqrt{3})$.
\\
The Einstein frame action takes the form 
\begin{eqnarray}\label{Ein_action}
S=\frac{1}{16\pi}\int d^4x \sqrt{-g^*}[R^*-2g^{*\mu \nu}\partial_\mu \varphi \partial_\nu\varphi - V(\varphi)] + S_{{\rm m}}
\end{eqnarray}
where $R^*$ is the Ricci scalar curvature with respect to the Einstein frame metric $g^{*\mu \nu}$ and $V(\varphi) = A^4(\varphi)U(\Phi(\varphi))$.
The explicit form of potential in the Einstein frame for R-squared gravity is $V(\varphi) = (1-exp(-2\varphi/\sqrt{3}))^2/4a$. In comparison to the Jordan frame, the Einstein frame's field equations are much simpler. The appearance of direct interaction between the scalar field $varphi$ and the matter fields in Einstein frame comes at a cost for simplifying the action and the field equations in this frame. By taking variation with respect to the metric $g^{*}_{\mu \nu}$ and the scalar field $\varphi$, the field equations in the Einstein frame are derived. The final quantities that we obtain have to be transformed back to the physical Jordan frame. In addition, the equation of state of the nuclear matter that we use will also be only in the Jordan frame \cite{Daniela1}.
\begin{eqnarray}\label{Ein_vary}
G^*_{\mu \nu}= 8\pi T^*_{\mu \nu} +  2\partial_\mu \varphi \partial_\nu\varphi - g^*_{\mu \nu}g^{*\alpha \beta} \partial_\alpha \varphi \partial_\beta\varphi - \frac{1}{2} V(\varphi)g^*_{\mu \nu}
\end{eqnarray}

\begin{eqnarray}\label{Ein_pot}
\nabla^*_{\mu} \nabla^{*\mu}\varphi - \frac{1}{4} \frac{dV(\varphi)}{d\varphi} = -4\pi \alpha(\varphi)T^*
\end{eqnarray}
where 
\begin{eqnarray}\label{Ein_alpha}
\alpha(\varphi) = \frac{d ln A(\varphi)}{d \varphi} = -\frac{1}{\sqrt{3}}
\end{eqnarray}
The Einstein frame energy-momentum tensor $T^*_{\mu \nu}$ is related to the Jordan frame energy-momentum tensor $T_{\mu \nu}$ via $T^*_{\mu \nu} = A^2 (\varphi)T_{\mu \nu}$ . In the case of a perfect fluid, the energy density, the pressure and the 4-velocity in the two frames are related via the formulae
\begin{eqnarray}\label{fluid_rel}
\rho_{*} = A^4(\varphi)\rho,
\quad p_{*} = A^4(\varphi)p,
\quad u^*_{\mu} = A^{-1}(\varphi)u_{\mu}.
\end{eqnarray}
The contracted Bianchi identities give the following conservation law for the Einstein frame energy-momentum tensor
\begin{eqnarray}\label{no}
\nabla^*_{\mu} T^{*\mu}_{\nu} =\alpha(\varphi)T^* \nabla^*_{\nu}\varphi
\end{eqnarray}
We will consider static and spherically symmetric spacetime described by the Einstein frame metric
\begin{eqnarray}\label{ }
ds^2_{*} =-e^{2\phi(r)}dt^2 + e^{2\lambda(r)}dr^2 + r^2(d\theta^2 + sin^2\theta d\vartheta^2)
\end{eqnarray}
We consider the matter source to be a perfect fluid and also that the perfect fluid and the scalar field are static and spherically symmetric. With these conditions imposed, the dimensionally reduced field equations are \cite {Daniela1, plb}
\begin{eqnarray}\label{tov1a}
\frac{1}{r^2}\frac{d}{dr}\left[r\left(1-e^{-2\lambda}\right)\right] = 8\pi \rho A^4(\varphi) + e^{-2\lambda}\left(\frac{d \varphi}{dr}\right)^2 + \frac{1}{2}V(\varphi)
\end{eqnarray}
\begin{eqnarray}\label{tov2a}
\frac{2}{r}e^{-2\lambda}\frac{d\phi}{dr} -  \frac{1}{r^2}\left(1-e^{-2\lambda}\right) = 8\pi p A^4 (\varphi) + e^{-2\lambda}\left(\frac{d \varphi}{dr}\right)^2 - \frac{1}{2}V(\varphi)
\end{eqnarray}
\begin{eqnarray}\label{tov3a}
\frac{d^2\varphi}{dr^2} + \left( \frac{d\phi}{dr} -\frac{d\lambda}{dr}+\frac{2}{r} \right)\frac{d\varphi}{dr} = 4\pi \alpha(\varphi) A^4 (\varphi)(\rho - 3p)e^{2\lambda} + \frac{1}{4}\frac{d V(\varphi)}{d\varphi}e^{2\lambda}
\end{eqnarray}
\begin{eqnarray}\label{tov4a}
\frac{dp}{dr} = -(p+\rho)\left[\frac{d\phi}{dr} + \alpha(\varphi)\frac{d \varphi}{dr}\right]
\end{eqnarray}
For the purposes of arriving at a numerical solution, the above equations can be further reduced to the following: 
\begin{eqnarray}\label{tov1b}
\frac{d \lambda}{dr} = e^{2\lambda}\left[4\pi \rho r A^4 + \frac{r e^{-2\lambda}}{2}\left(\frac{d \varphi}{dr}\right)^2 + \frac{r(1-A^2)^2}{16a} - \frac{(1-e^{-2\lambda})}{2r}\right]
\end{eqnarray}

\begin{eqnarray}\label{tov2b}
\frac{d \phi}{dr} = e^{2\lambda}\left[4\pi p r A^4 + \frac{r e^{-2\lambda}}{2}\left(\frac{d \varphi}{dr}\right)^2 - \frac{r(1-A^2)^2}{16a} + \frac{(1-e^{-2\lambda})}{2r}\right]
\end{eqnarray}

\begin{eqnarray}\label{tov3b}
\frac{d^2 \varphi}{dr^2} = e^{2\lambda}\left[ \frac{A^2(1-A^2)}{4\sqrt{3}a}- \frac{4 \pi A^4 (\rho - 3 p)}{\sqrt{3}}\right] - \frac{d \varphi}{dr}\left(\frac{d \phi}{dr} - \frac{d \lambda}{dr} + \frac{2}{r}\right)
\end{eqnarray}

\begin{eqnarray}\label{tov4b}
\frac{dp}{dr} = -(p+\rho)\left[\frac{d\psi}{dr} - \frac{1}{\sqrt{3}}\left(\frac{d \varphi}{dr}\right)\right]
\end{eqnarray}
The Jordan frame quantities $\rho$ and p are naturally connected via the EoS of the neutron star matter $p = p(\rho)$. Furthermore, we should impose the standard boundary conditions – regularity at the star's center and asymptotic flatness at infinity. The equations describing the spacetime metric and the scalar field outside the neutron star are obtained by setting $\rho = p = 0$.
In order to solve our systems of differential equations for the interior and the exterior of the neutron star, we should provide the EoS for the neutron star matter $p = p(\rho)$ and impose the boundary conditions. 
We solve the interior and the exterior problems together using the natural Einstein frame boundary conditions at the center of the star which are as follows:
\begin{eqnarray}\label{bc}
\rho(0) = \rho_c, \quad \lambda(0) = 0, \quad \frac{d\varphi}{dr}(0) = 0 
\end{eqnarray}
At infinity,
\begin{eqnarray}\label{bc_inf}
\lim\limits_{r\to\infty} \phi(r) = 0, \quad \lim\limits_{r\to\infty} \varphi(r) = 0
\end{eqnarray}
To get the neutron star mass and radius, we need to solve numerically the equations (\ref{tov1b}, \ref{tov2b} and \ref{tov4b}) together with the boundary conditions equations (\ref{bc}) - (\ref{bc_inf}) from center to surface of the star and then surface to infinity. In order to integrate those coupled differential equations from center of the star, we need to fix the value of $\phi$ for each central density of the star at the center such that it should exponentially decay and go to zero at infinity. The behaviour of $\phi$ with integrating radius can be seen in fig 5 of Ref. \cite{plb}. %PLB Paper
The radius of the star is calculated from the requirement of pressure vanishing at the stellar surface and the mass is taken from the asymptotic expansion of the metric functions at infinity.  In Einstein frame, the star mass $m(r)$ can be defined according to the relation $e^{-2 \lambda}=1-\frac{2 m}{r}$. The requirement of asymptotic flatness fixes the constraints on the scalar curvature and the mass parameter as, $$\lim _{r \rightarrow \infty} R(r)=0, \quad \lim _{r \rightarrow \infty} m(r)=0$$  It is important to note that for the considered R-squared gravity the mass in the Einstein and the Jordan frame coincide, while the physical Jordan frame radius of the star $R_S$ is connected to the Einstein frame one $r_s$ through the following relation 
\begin{eqnarray}\label{Rs}
R_S = A(\varphi(r_s))r_s
\end{eqnarray}
The dimensions of the parameter $a$ is in terms of $r^2_g$, where $r_g = 1.47664$ km corresponds to one solar mass. The $m(r)$ profile for inside and outside star for both the Einstein and the Jordan frame can be seen in fig 4 of Ref. \cite{plb}. %PLB Paper
\subsection{Tidal Love number}
Consider a static, spherically symmetric star in R-squared gravity. The stationary perturbations of the metric can be separated into magnetic/axial and electric/polar types. In this study, we will calculate only the tidal Love number for the polar type and its corresponding tidal deformability. Even though the absolute value of the axial Tidal Love Number can increase dramatically in R-squared gravity\cite{Daniela2}, recent estimates indicate that they would have no observable impact on the gravitational wave signal, even for the next generation of gravitational wave detectors \cite{GWdetector}. 

\subsubsection{Polar type}
The electric/polar perturbations of the perturbed Einstein frame metric in the Regge-Wheeler gauge can be written in the form\cite{RW}    

\begin{eqnarray}\label{axial_mat}
H^{polar}_{\mu\nu} = \begin{bmatrix}
-e^{2\psi_0}H_0(r) & H_1(r) & 0 &0\\
H_1(r) &e^{2\lambda_0}H_2(r) &0 &0\\
0 &0 &K(r)r^2 &0\\
0 &0 &0 &K(r)r^2 sin^2\theta
\end{bmatrix} \times  Y_{lm}(\theta, \phi)
\end{eqnarray}
where $Y_{lm}(\theta, \phi)$ are the spherical harmonics. The perturbations of the scalar field, energy density and the pressure can be decomposed into $\delta\varphi = \delta \tilde{\varphi}(r) Y_{lm}(\theta, \phi)$, $\delta\rho^* = \delta \tilde{\rho}(r) Y_{lm}(\theta, \phi)$, $\delta p^* = \delta \tilde{p}(r) Y_{lm}(\theta, \phi)$. 

After perturbing the Einstein frame field equations of the $f(R)$ gravity coupled to a perfect fluid, it can be shown that $H_0 = -H_2$ and $H_1 = 0$ and $\delta \tilde{\rho}(r)$, $\delta \tilde{p}(r)$, and K can be written as a function of $H_0$ and $\tilde{\varphi}$.  We also obtain the following two equations for $H_0= -H_2 = H$ and $\tilde{\varphi}$ governing the stationary perturbations of the static and spherically symmetric stars in $f(R)$ gravity: 

\begin{eqnarray}\label{d2H}
%\begin{multlined}
&&\frac{d^2H}{dr^2} +\Bigg\{\frac{2}{r} + e^{2\lambda_0} \left[ \frac{1-e^{-2\lambda_0}}{r} + 4\pi r(p^*_0 - \rho^*_0) - \frac{r}{2}V(\varphi_0)    \right] \Bigg\}\frac{dH}{dr} \nonumber  \\ &+&e^{2\lambda_0}\Bigg\{-\frac{l(l+1)}{r^2} + 4\pi \left[9p^*_0 +5\rho^*_0 + \frac{\rho^*_0 + p^*_0}{\tilde{c}^2_s}  - \frac{1}{4\pi}V(\varphi_0) \right] - 4\psi^{'2}_0 e^{-2\lambda_0} \Bigg\} H  \\
&+& e^{2\lambda_0} \Bigg\{-4\varphi'_0 r \left[ \frac{1-e^{-2\lambda_0}}{r} + 8\pi p^*_0 + e^{-2\lambda_0} \varphi^{'2}_0 -\frac{1}{2}V(\varphi_0) \right] - \frac{16\pi}{\sqrt{3}}\left[ (\rho^*_0 - 3p^*_0) + (\rho^*_0 + p^*_0)\frac{1-3\tilde{c}^2_s}{2\tilde{c}^2_s} + V^{'}(\varphi_0)\right] \Bigg\}\delta\tilde{\varphi} = 0 \nonumber
%\end{multlined}
\end{eqnarray}

\begin{eqnarray}\label{d2delphi}
%\begin{multlined}
&&\frac{d^2\delta\tilde{\varphi}}{dr^2} + \left(\psi^{'}_0 - \lambda^{'}_0 +\frac{2}{r}\right)\frac{d\delta\tilde{\varphi}}{dr} -  e^{2\lambda_0}\Bigg\{ \frac{l(l+1)}{r^2}+e^{-2\lambda_0}4\varphi^{'2}_0 + \frac{1}{4}V^{''}(\varphi_0) - \frac{8\pi}{3}\left[-2(\rho^*_0 - 3p^*_0) + (\rho^*_0 + p^*_0)\frac{1-3\tilde{c}^2_s}{2\tilde{c}^2_s} \right] \Bigg\}\delta\tilde{\varphi} \nonumber \\ &+&e^{2\lambda_0} \Bigg\{-2 \psi^{'}_0 \varphi^{'}_0 e^{-2\lambda_0} - \frac{4\pi}{\sqrt{3}}\left[(\rho^*_0 - 3p^*_0) + (\rho^*_0 + p^*_0)\frac{1-3\tilde{c}^2_s}{2\tilde{c}^2_s} \right] + \frac{1}{4}V^{'}_0(\varphi_0)\Bigg\}H = 0
%\end{multlined}
\end{eqnarray}
Here $\lambda_0$, $\psi_0$, $\varphi_0$, $p_0$ and $\rho_0$ are the corresponding unperturbed variables taken from the background neutron star solutions and $\tilde{c}_s$ is the Jordan frame speed of sound defined as $\tilde{c}^2_s = \partial p/\partial \rho$.

As discussed above, the scalar field mass of $f(R)$ gravity is nonzero which implies that both the background scalar field $\varphi_0$ and its perturbation $\delta\tilde{\varphi}$ drop off exponentially at infinity.  This implies that the corresponding scalar field tidal Love number is zero.
The asymptotic behaviour of $H$ at large r is
\begin{eqnarray}\label{Hasymptote}
H = \frac{c_1}{r^{l+1}} + \mathcal{O}\left(\frac{1}{r^{l+2}}\right) + c_2r^l + \mathcal{O} (r^{l-1}).
\end{eqnarray}
The tidal Love number $k_2$ is connected to the coefficients $c_1$ and $c_2$ in the above equation as follows: 
\begin{eqnarray}\label{k2_th}
k^{polar}_l = \frac{1}{2R^{2l+1}_S} \frac{c_1}{c_2}
\end{eqnarray}
In pure GR, the ratio $c_1/c_2$ is usually determined by matching at the stellar surface the numerical solution of H from the interior of the neutron star to the analytical solution exterior to it. This approach cannot be applied to the case here since the equations for $H$ and $\delta\tilde{\varphi}$ are coupled and, in general, there exists no analytical solution outside the star. Since we know that the scalar field is massive, the problem is simplified. Since the scalar field and its perturbation die out exponentially at distances larger than the Compton wavelength of the scalar field $\lambda_{\varphi} = 2\pi/m_{\varphi} = 2\pi\sqrt{6a}$, the analytical solution of GR can be used. The inner and outer solutions are matched at a radius $r_{match}$ where the scalar field and its perturbation are negligible rather than at the stellar surface. Since $r_{match}$ is connected to the Compton wavelength of the scalar field, $r_{match}$ is not a constant but
a function of the parameter a.
For $l=2$, we obtain
\begin{eqnarray}\label{polark2}
\begin{multlined}
k^{polar}_2 = \frac{8C^5_1}{5} (1-2C)^2 \left[2+2C(y-1)-y\right] \times \\ \Big\{2C\left[6-3y+3C(5y-8)\right] +4C^3\left[13-11y+C(3y-2)+2C^2(1+y)\right] + \\ 3(1-2C)^2\left[2-y+2C(y-1)\right] log(1-2C) \Big\}^{-1}
\end{multlined}
\end{eqnarray}
Here $y = rH^{'}/H$, $C_1= M/R_S $ is the compactness of the star and $C=M/r_{match}$. The value of $y$ is calculated from the numerical solution of equations 28 and 29 from $r = 0$ to $r = r_{match}$.

To compute this polar tidal love number numerically, we need to solve the simultaneous second order differential equations 28 and 29 with some initial values for $H$, $dH/dr$. $\delta\varphi$ and $d \delta\varphi/dr$ at the center of the neutron star. We have used $\chi^2$ minimization method to arrive at values for $\delta\varphi$ at the center of the star such that the perturbation exponentially decays to the neighborhood of zero around the Compton wavelength.

For the values of $a$, it is only reasonable to assume that the scalar field's Compton wavelength to be smaller than the orbital separation between merging neutron stars at the time when the ground-based detectors are sensitive to observe and calculate the tidal deformability. Assuming that we can detect the emitted gravitational waves with current instruments when the orbital separation decreases to a few hundreds of kilometers, we chose $a \leq 100$ which gives a Compton wavelength $\leq$ 154 km.

\section{Results}
\begin{figure*}[t]
  \includegraphics[height=6.5 cm ]{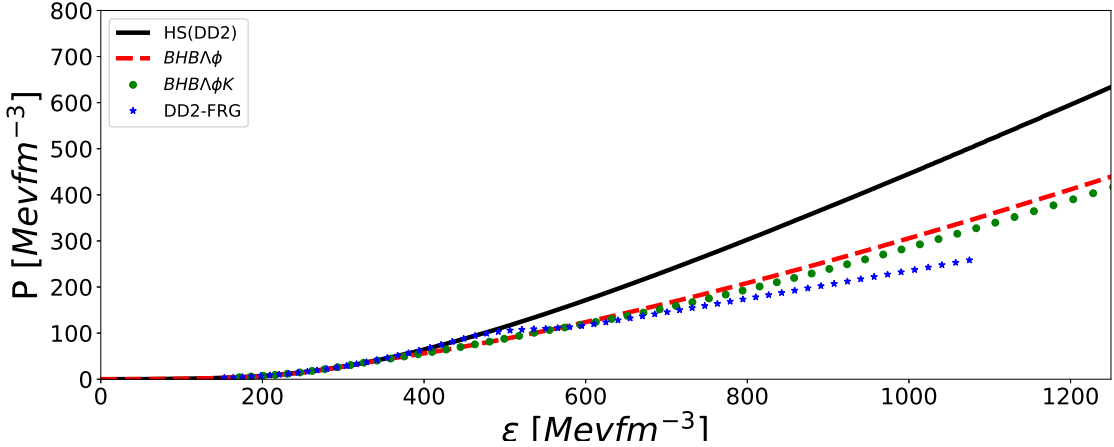}
  \caption{Neutron star EoSs (energy density ($\epsilon$) versus pressure (P)) for different models - both nucleon only and with additional exotic particles such as ${\bf \Lambda}$, $K^-$ condensates and quarks. }
  \label{fig:eos}
  \end{figure*}
\subsection{Neutron Star Model EoS}

The motivation of this work is to investigate the effect of strangeness on the properties of NS in modified gravity. The interior temperature of the NS is low  compared to the Fermi energy of the constituent fermions. Hence the dense matter relevant to the NS core can be modeled in terms of zero temperature EoS. We consider EoS corresponding to dense matter with A) nucleons only B) nucleons with additional strange components.  All the models considered in this work obey the mass-radius constraints of astrophysical observations \cite{fonseca2016nanograv, antoniadis2013t, demorest5788two}.

{\bf Nuclear EoS}

Ideally, the dense matter in the NS core consists of neutron, protons, and electrons and is represented by nucleonic EoS. We consider 
%two such EoS here i)Sly4: a soft EoS, based on the many-body calculations using single effective nuclear Hamiltonian, ii) 
HS(DD2) EoS, an EoS based on relativistic mean field model with density-dependent parameter set \cite{hempel2010statistical, char2014massive}. 
\iffalse
The Sly4 EoS of neutron star matter, based on the effective nuclear interaction SLy(Skyrme Lyon)\cite{Dou2001} is one of the few  models of unified EoS, where the crust and core segments are obtained using same physical model. 
For Sly4 EoS the transitions between the outer crust and the inner crust, and between the inner crust and the core are obtained as a result of many-body calculation. A proper core-crust matching is very crucial to avoid uncertainties in the macroscopic properties of the stars as emphasised in Ref \cite{fortin2016neutron}.
%The maximum mass of static neutron stars is  $2.05 M_{solar}$ and the corresponding radius is 9.99 km in this model.
%%A particular reason for choosing this nucleonic EoS is to compare our results with the previously published ones\cite{}. 
\fi 
HS(DD2) is unified EoS in the sense that it employs the same Lagrangian density to describe low-density crust as well as the high-density core and allows a smooth transition between the core and crust. The  non-homogeneous matter consists of an ensemble of nuclei and nucleons and is described within the framework of the nuclear statistical equilibrium (NSE) model whereas the strong nucleon-nucleon interaction in the uniform dense nuclear matter part  is mediated by exchanges of scalar $\sigma$, vector $\omega$ and $\rho$ mesons \cite{hempel2010statistical,rmp}.
The high-density behaviour is considered through the density dependence of the meson-baryon couplings. The saturation density, mass of $\sigma$ meson, couplings at the saturation density, obtained by fitting properties of finite nuclei is known as the DD2 parameter set. Nuclear matter properties at the saturation density are compatible with those obtained from nuclear physics experiments \cite{rmp}.  

\begin{figure*}[t]
  \includegraphics[ height=11 cm]{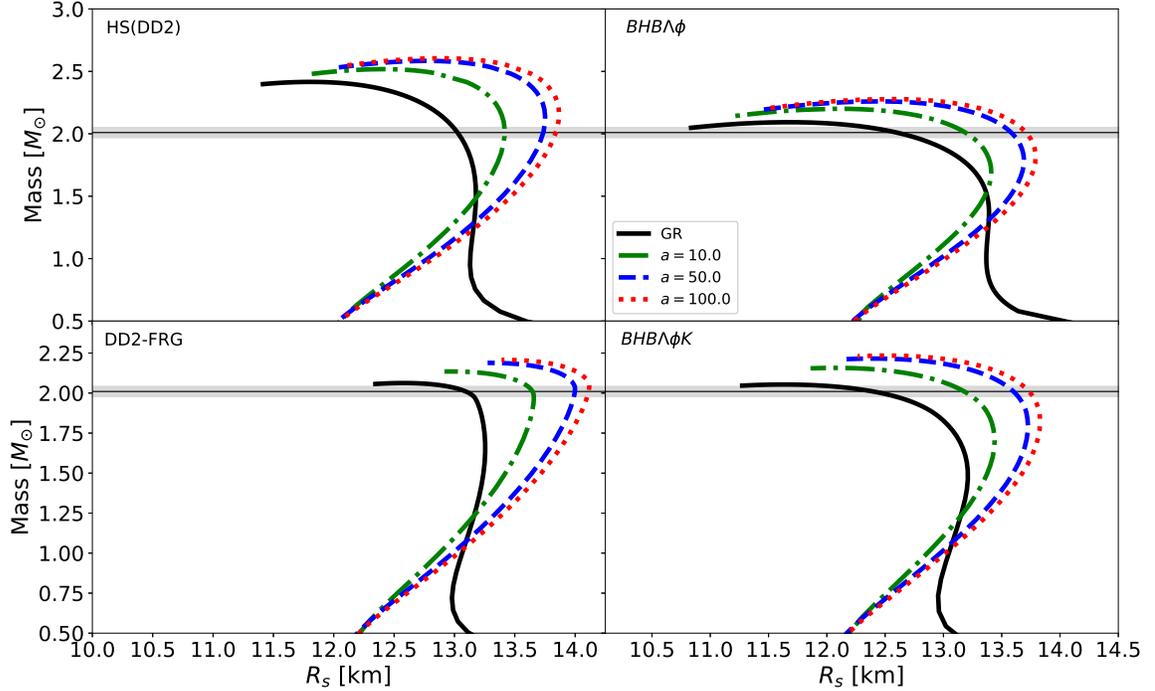}
  \caption{Mass-radius relation of EoS under study in GR and f(R) for a range of parameter "a". The gray band in the figures represent the observational constraints on mass\cite{antoniadis2013t,demorest5788two}}
  \label{fig:MRall}
  \end{figure*}
 
 %\begin{figure*}[t]
 % \includegraphics[ height=11 cm]{SLY4_all.eps}
 % \caption{SLY4 EoS and various properties of the neutron stars such as Mass-Radius, $k_2$, $\Lambda$}
 % \label{fig:Sly4}
 % \end{figure*}

\begin{figure*}[t]
  \includegraphics[ height=11 cm]{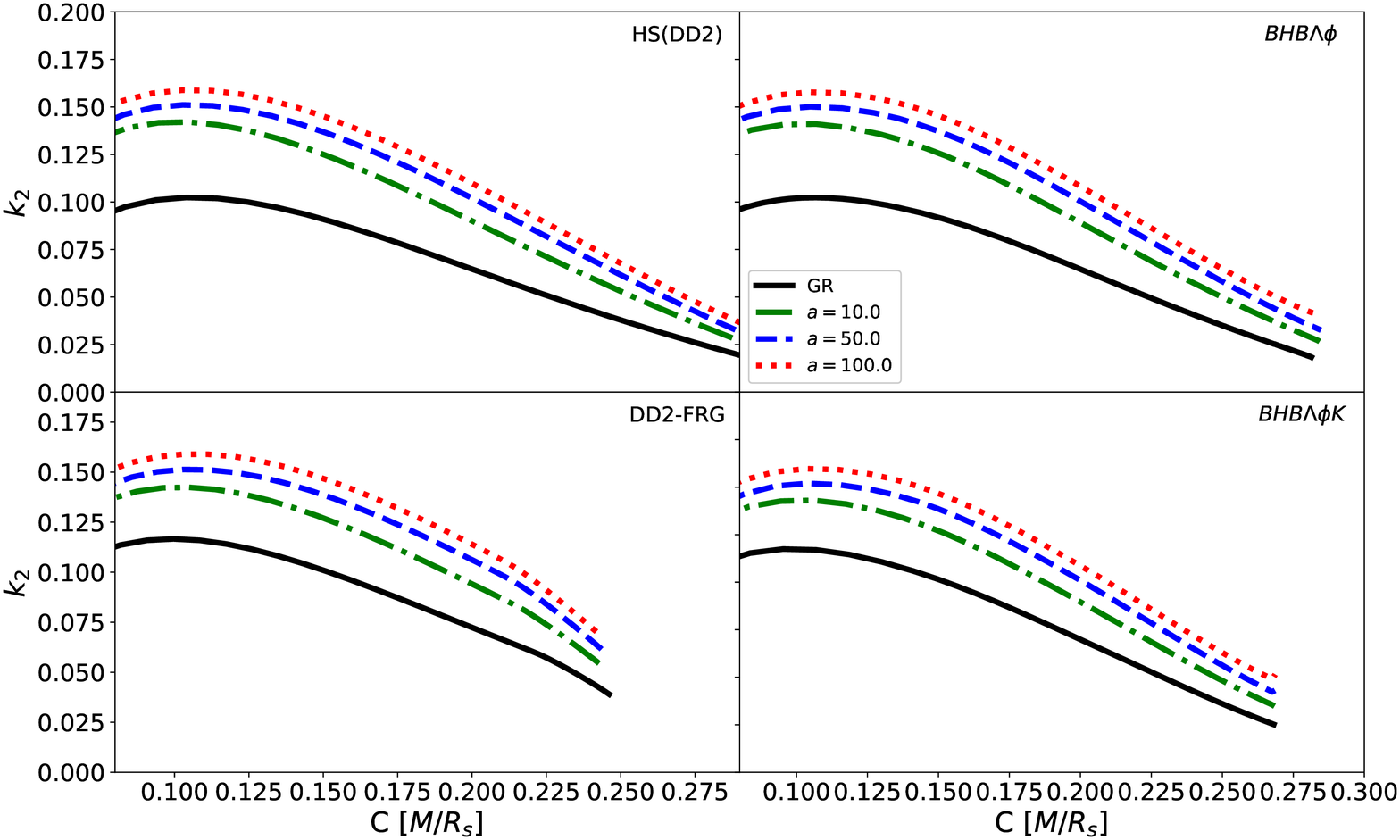}
  \caption{Tidal Love Number ($k_2$) versus Compactness (C) for density-dependent EoS under study }
  \label{fig:k2}
  \end{figure*}

\begin{figure*}[t]
  \includegraphics[ height=11 cm]{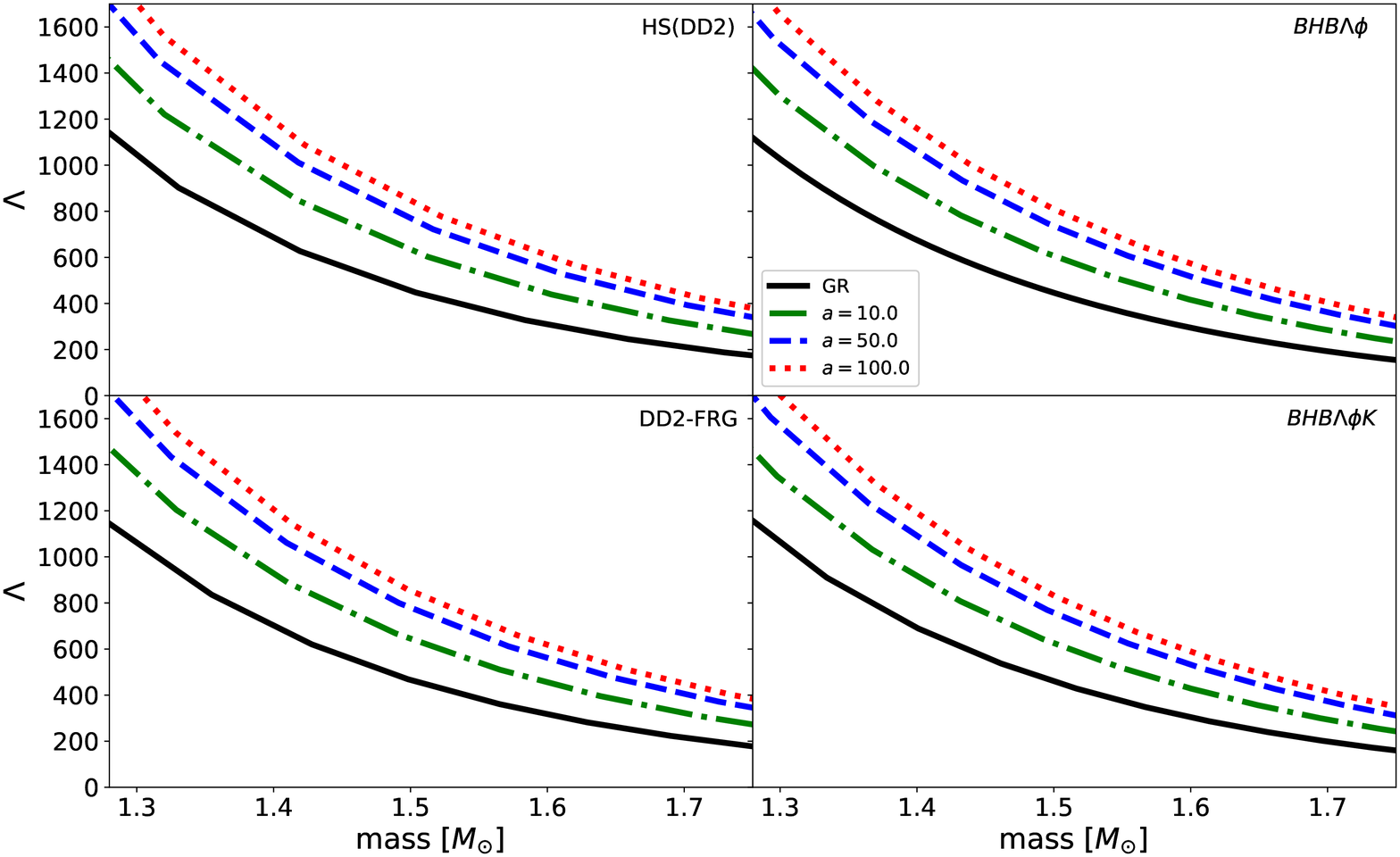}
  \caption{The neutron star tidal deformability ($\Lambda$) versus mass for the EoSs under study }
  \label{fig:lambda}
  \end{figure*}

{\bf Strange EoS}

The high-density core of the NS is hypothesized to be populated with strange hadrons such as antikaon condensates, $\bf \Lambda$ hyperons and even quark matter under tremendous pressure.  The onset of a new degree of freedom softens the EoS and supports a star that is less massive and less compact compared to the ones comprising of only nucleons. We consider three strange EoS consisting of i) ${\bf \Lambda}$ hyperons (BHB$\Lambda \phi$) \cite{banik2014new}, ii) ${\bf \Lambda}$ hyperons  and $K^-$ mesons (BHB$\Lambda \phi K^-$) \cite{apj2021} and iii) 2 flavor quark matter (DD2-FRG)  \cite{oosA} in addition to the nucleons. All three hybrid EoS are constructed from the HS(DD2) EoS for hadronic matter.

The ${\bf \Lambda}$ hyperon is populated when its in-medium chemical potential matches that of neutron. The density-dependent ${\bf \Lambda}$-vector meson hyperon couplings are obtained from the SU(6) symmetry of the quark model \cite{dov,sch96} i.e.  $g_{\omega {\bf \Lambda}} = \frac{2}{3} g_{\omega N}, \quad g_{\rho {\bf  \Lambda}} = 0, \quad g_{\phi {\bf \Lambda}} = -\frac{\sqrt 2}{3} g_{\omega N}.$  On the other hand, the scalar meson coupling to {\bf $\Lambda$} hyperons  are determined from their potential depth (-30MeV) in normal nuclear matter, obtained from the experimental data of the single particle spectra of ${\bf \Lambda}$ hypernuclei. Other baryons of the octet are not considered mainly due to the uncertainty in their experimental data. This EoS is known as BHB$\Lambda \phi$  EoS.
 
In the BHB$\Lambda \phi K^-$ EoS, the kaon-nucleon interaction is considered in the same footing as the nucleon-nucleon interaction.
The $K^-$ condensate appears in the system when its in-medium energy equals to its chemical potential given by $\mu_{K^{-}}=\mu_n-\mu_p$. The kaon-meson couplings do not depend on density and are constants. They are estimated exploiting the quark model and iso-spin counting rule, i.e., $g_{\omega K} = \frac{1}{3} g_{\omega N}$ and $g_{\rho K} = g_{\rho N}$ \cite{banik01,sch96}. The scalar coupling constant is determined from the real part of $K^{-}$ optical potential at the saturation density, which we take as -120 MeV.

Finally, we consider DD2-FRG EoS for quark matter constructed within a model based on the underlying chiral symmetry breaking of QCD. This is a non-perturbative two- and three-flavored quark model  obtained with the  functional renormalization group (FRG) \cite{oosA}. The maximum hybrid star mass of about $2.1 M_{\odot}$ for the two-flavor model complies well with current mass observations, whereas the strange quark model does not satisfy the mass limit. So in our calculation we used the two-flavored DD2-FRG EoS. For the HS(DD2) EoS, the GW170817 tidal deformability is $\tilde\Lambda \sim 795$ for a mass ratio of 0.8 of the two coalescing stars. This value does not change significantly for both the hybrid EoS  as the quark matter onset occurs only for masses slightly below $1.8 M_{\odot}$ and higher.

The above-mentioned EoSs are plotted in fig. \ref{fig:eos}. The nucleonic HS(DD2) EoS is denoted by solid black line. The other curves represent EoS containing additional exotic particles such as {\bf $\Lambda$} hyperons (BHB$\Lambda \phi$), {\bf $\Lambda$} hyperons and anti(kaon) condensates (BHB$\Lambda \phi K^-$) and quarks (DD2-FRG) in the HS(DD2) model. The addition of strange particles softens the EoS. The DD2-FRG is evidently the softest of the lot.
%\section{Results}

For GR, we derive and solve the TOV equations for a spherically symmetric, non-rotating neutron star to obtain the mass and radius given an EoS and appropriate boundary conditions. For a chosen central pressure P(0) and $\epsilon(0)$, we integrate from the center to the surface of NS with boundary conditions at the center [m(0)=0] and on the surface [P($R_s$)=0], total mass [M=m($R_s$)] enclosed in the star of radius $R_s$. In fig. \ref{fig:MRall}, we represent these using black solid lines. The four panels are for the four RMF EoS, mentioned above. A comparison of these four mass-radius curves clearly shows the effect of strangeness. Softer the EoS, lower is the maximum mass, it can support. All of them are above $2M_{\odot}$, the highest being $2.41M_{\odot}$ for HS(DD2). The gray band in the figures represent the observational constraints \cite{antoniadis2013t}. The radius corresponding to the maximum masses in GR is in the range of 9.92 to 12.56 km.   

Spherically symmetric configurations are obtained by solving the modified TOV equations [\ref{tov1b}-\ref{tov4b}] for the R-squared gravity model in non-perturbative formalism for the same set of realistic EoS. The numerical integration of the coupled differential equations are done using RK4 method with a step size of 0.001 km and employing spline interpolation to obtain the intermediate energy densities. The mass and corresponding radius for each entry of the EoS are calculated to plot the corresponding mass-radius relations. This procedure is repeated for the value of $a =$ 10, 50, and 100 and the different EoS used in this study. For $a =$  10, the radius corresponding to the maximum mass is in the range of $10.75 - 12.97$ km; for $a =$ 50, it is in the range of $11.06 - 13.27$ km; for $a =$ 100, it is in the range of $11.13 - 13.39$ km.
%In figure 3, we plot the mass vs radius for the before-mentioned EoS, both nucleonic and strange. .......Name ...and panels.....black solid line is for GR,....... ... The maximum mass condition is met for all these EoS for $a=0$ (GR) and 10, 50, 100; and we can see that 
For each EoS as $a$ increases the mass and radius of the neutron star increase compared to the GR mass and radius of the corresponding EoS. As evident from the table \ref{tab:tab1}, the radius of NS at 1.4 $M_{\odot}$ ($M_{1.4}$) for all $a$ is also well within the bounds of NICER results ($13.02^{+1.24}_{-1.06}$) \cite{Miller_2019}. For the four EoS considered in this study, the change in maximum mass of NS with $a =$ 10 from the GR maximum mass is from $0.072 - 0.109$ $M_{\odot}$ while the increase in radius for $M_{1.4}$ is within $0.12 - 0.31$ km. For $a =$ 50, the mass varies from $0.126 - 0.174$ $M_{\odot}$, while the increase in radius for $M_{1.4}$ is within $0.32 - 0.54$ km; and for $a =$ 100, it varies from $0.144 - 0.196$ $M_{\odot}$, while the increase in radius for $M_{1.4}$ is within $0.39 - 0.61$ km. 
Among the EoS, we see that HS(DD2) has the maximum mass both in GR and $f(R)$. In GR, addition of ${\bf \Lambda}$ hyperon, $K^-$, and quarks to the EoS reduces the maximum mass considerably. Similar behaviour is seen for different values of $a$ in $f(R)$.

%In figure 1, we plot the energy vs pressure of the following EoS: HS-DD2, BHB$\Lambda\phi$, DD2-FRG, BHB$\Lambda\phi K$, and SLY4. We have taken both soft and stiff EoS

\begin{table}
\caption{\label{tab:tab1}%
Data table of maximum mass ($M_{max}$), Radius at 1.4 $M_\odot$ ($R_{S,1.4}$), and $\Lambda$ at 1.4$M_\odot$ ($\Lambda_{1.4}$). 
}
\begin{center}
\begin{tabular}{ |p{1.6cm}||p{0.55cm}|p{0.55cm}|p{0.55cm}|p{0.55cm}|| p{0.7cm}|p{0.7cm}|p{0.7cm}|p{0.7cm}||p{0.9cm}|p{0.9cm}|p{1.0cm}|p{1.0cm}|| }
 \hline
\multicolumn{1}{|c|}{}&\multicolumn{4}{|c|}{$M_{max}$ ($M_\odot$)}&\multicolumn{4}{|c|}{$R_{S,1.4}(km)$}&\multicolumn{4}{|c|}{$\Lambda_{1.4}$} \\
 \hline
 a($r^2_g$) &0&10&50&100&0&10&50&100&0&10&50&100\\
 \hline
 HS(DD2)& 2.41& 2.52& 2.59 & 2.61&13.10&13.31&13.53&13.59&627.05&846.59&1011.52&1087.11\\
 BHB$\Lambda\phi$&2.09&2.20&2.26&2.28&13.04&13.24&13.45&13.64&648.33&781.99&933.53&1003.85\\
 DD2-FRG&2.06&2.13&2.19&2.21&13.23&13.34&13.54&13.62&620.22&888.47&1061.15&1139.76\\
 BHB$\Lambda\phi$K &2.05&2.16&2.22&2.23&13.20&13.33&13.54&13.62&689.27&807.68&964.54&1036.10\\
 %SLY4&2.05&2.14&2.19&2.21&11.61&11.92&12.14&12.22&286.87&432.19&581.38&661.02\\
 \hline
\end{tabular}
   \end{center}
\end{table}

\iffalse
Fig. \ref{fig:Sly4} is presented as a control to compare our numerical analysis with the existing literature \cite{Daniela1}. In the first panel, we plot the energy versus pressure of SLY4 EoS, which is a soft nucleonic EoS. In the second panel, we plot the corresponding mass versus radius. The black line represents the TOV results for GR case, which is within the mass radius constraints given by observations \cite{antoniadis2013t, demorest5788two}. The dotted lines represent $f(R)$ results for $a =$ 10, 50, and 100. Clearly, they all satisfy the mass and radius constraints. In the third panel, we plot $k_2$ polar versus compactness C. \cite{Daniela1} For GR, their value of maximum $k_2$ 0.1025 is at C = 0.10, we have our $k_2$ maximum of 0.107 at C = 0.103. For $a =$ 50, their value of maximum $k_2$ 0.115 is at C = 0.10, we have our $k_2$ maximum of 0.1446 at C = 0.124. In addition to this, we have also calculated the tidal deformability ($\Lambda$). In the fourth panel, we plot $\Lambda$ versus mass.
\fi
In fig. \ref{fig:k2}, we plot $k_2$ versus C (dimensionless compactness) for each EoS. The plots indicate that as $a$ increases the $k_2$ value corresponding to the compactness increases significantly compared to values in GR. For a fixed EoS the value of $k_2$, compared to the GR value, can vary from $24-62{\%}$ for the considered range of values of $a$. 
%The value of $k_2$ at $1.4 M_{\odot}$for GR
%While the nucleonic EoS shows the maximum variance, the EoS with strange particles/quarks show the least variance as a function of $a$.
In the case of GR, we see that the value of $k_2$ is in the range of $0.089 - 0.098$. However, as $a$ increases, we see that for $a =$ 10, $k_2$ ranges from $0.120 - 0.124$, which is a $24 - 37 {\%}$ increase; for $a = $50, $k_2$ ranges from $0.133 - 0.137$, which is a $38 - 52 {\%}$ increase; for $a = $100, $k_2$ ranges from $0.142 - 0.145$, which is a $46 - 62 {\%}$ increase. Among the EoS in GR, the addition of ${\bf \Lambda}$ hyperon, $K^-$, and quarks to the EoS increases the value of $k_2$ at $R_{S, 1.4}$. However, in $f(R)$, while the addition of quarks to the EoS increases the value of $k_2$ at $R_{S, 1.4}$, the addition of ${\bf \Lambda}$ hyperon and $K^-$ decreases the value of $k_2$ at $R_{S, 1.4}$ for different values of $a$ in $f(R)$.

In fig. \ref{fig:lambda}, we plot tidal deformability versus mass for each EoS. The plots indicate that as $a$ increases the value of $\Lambda$ corresponding to the mass of NS increases with respect to the corresponding values in GR. The change in $\Lambda_{1.4}$ of NS with $a =$ 10 with respect to the GR $\Lambda_{1.4}$ is from $118.41 - 268.25$, which corresponds to a increase of $17-43{\%}$ increase. For $a =$ 50, the $\Lambda_{1.4}$ varies from $275.27 - 440.94$ which corresponds to a increase of $40-71{\%}$ increase; and for $a =$ 100, it varies from $346.83 - 519.54$ which corresponds to a increase of $50-84{\%}$ increase. While all $a$ up to 100 were allowed by the constraints on mass and radius, we see that GW170817 constraint on the tidal deformability implies that value of $a$ should be less than 10 for all EoSs.  

In the case of GR, the addition of quark matter reduces the value of $\Lambda$; while in $f(R)$, the value increases compared to nucleons only EoS. The quark EoS has the highest values of $\Lambda$ among the EoS in $f(R)$. In GR, addition of kaons and hyperons increases the value of $\Lambda$; whereas in $f(R)$, the value decreases than in nucleons only EoS for different values of $a$. 
%Comparing with the softer nucleonic EoS, the value of $\Lambda$ is about two times higher both in GR and f(R).
%rana paper...values for lambda.

%It has been shown \cite{Arapo_lu_2011} that observational constraints of 2$M_{\odot}$ from the PSR J1614-2230 excludes many of the possible soft EoS if one presumes GR as the only classical theory of gravity. 

\section{Conclusion}
In the recent years, there is an increasing interest in studying modified gravity theories, mainly to understand the universe's rapid expansion, which has been supported by many independent observations \cite{perlmutter1999measurements}. The modified theories of gravity are well motivated and compatible with observational evidence without requiring the addition of an ad hoc component of matter, that have so far eluded discovery. However, such modifications should pass all astrophysical checks before becoming approved or rejected. In the weak regime, the solar system test and laboratory experiments are used to verify the validity of the majority of these models. However, compact stars, such as neutron stars, form excellent models for testing the impact of modified gravity in the strong regime \cite{ozel2016masses}.

The investigation of the EoS at supra-nuclear densities is still an open problem as we do not know the nature of nuclear interactions at these densities. One approach employed is to try different nuclear models and construct the dense matter EoS relevant to NS. The EoS is then used to to calculate various structure properties such as mass, radius, tidal deformability, moment of inertia, etc. in GR. The EoS that satisfy the astrophysical observations are accepted. The other approach is to modify GR to see its effect on the NS properties. Currently, the difference between the R-squared gravity and GR overlap with the uncertainties in the EoS which clouds our understanding \cite{Daniela1}. Any information that aids in constraining the free parameter '$a$' in R-squared gravity will help us better understand the above-mentioned uncertainties.

In this work, we used the non-perturbative approach in R-squared gravity to derive the modified TOV equations to study spherically symmetric, non-rotating neutron stars \cite{Daniela1}. For this, we consider relativistic mean field EoS with density-dependent couplings [HS(DD2)] model EoS. As our objective is to study the effect of exotic particles such as (${\bf \Lambda}$, $K^-$, and quark) in modified gravity, we consider EoS based on DD2 model. The nucleonic as well as the exotic EoS satisfy the maximum mass constraints of NS. We explored the mass versus radius, compactness versus $k_2$, and mass versus tidal deformability relations for EoS. We found that while the addition of exotic particles reduces the maximum mass for all $a >$ 0, the addition of ${\bf \Lambda}$ and $K^-$ reduces the $\Lambda_{1.4}$ for $a >$ 0, whereas, the addition of quarks increases the $\Lambda_{1.4}$. The maximum-mass constraint allows a wide range of values for $a$ (from 0-100) in R-squared gravity. To improve upon this, we further investigated the allowed range of values for $a$ using the tidal deformability in the light of results from GW170817 \cite{abbott2018gw170817}.

We are able to constrain the value of $a$ to be less than 10 for the type of EoS considered in this study. We have confined our attention to the EoS based on density-dependent relativistic mean field model with nucleonic as well as exotic degrees of freedom. For any given $a > 0$, the tidal deformability in f(R) is always greater than that in GR. Some very soft EoS may satisfy the observational limits of tidal deformability in f(R), but not in GR. However, they can never meet the maximum mass criteria and therefore will be ruled out. In future, we would like to extend this work in a model-independent approach by considering different family of EoSs to study various  universal relations, such as the “I-Love-Q” and “C-Love” relations \cite{Chan:2014kua,Yagi:2013awa} in the framework of $f(R)$ Gravity, which is already well explored in GR.

\section{Acknowledgements}
Authors T.M and S.B acknowledge the DAE-BRNS grant received under the BRNS project No.37(3)/14/12/2018-BRNS. The authors would also like to thank Prasanta Char for useful discussions. 


\end{document}